\documentclass[twocolumn,showpacs,aps,prl,superscriptaddress]{revtex4}

\usepackage{graphicx}
\usepackage{dcolumn}
\usepackage{amsmath}
\usepackage{epsfig}

\input babarsym

\def\lumi    {$84.4\pm 0.9$ million}

\def\luminapprox   {$84$ million}
\def\onlumi {\ensuremath 78} 
\def\limitrhoo {\ensuremath 1.2\times 10^{-6}}
\def\limitrhop {\ensuremath 2.1\times 10^{-6}}
\def\limitomega {\ensuremath 1.0\times 10^{-6}}
\def\limitgeneric {\ensuremath 1.9\times 10^{-6}}
\def\limitratio {\ensuremath 0.047}

\def\Bp {B^+}

\def\Bpm{B^{\pm}}
\def\bsg    {\ensuremath {b \to s \gamma}}
\def\bdg    {\ensuremath {b \to d \gamma}}
\def\bkg    {\ensuremath {\B \to \Kstar \gamma}}
\def\bkog    {\ensuremath {\Bz \to \Kstarz \gamma}}

\def\bkpg    {\ensuremath {\Bp \to \Kstarp \gamma}}

\def\brg    {\ensuremath {B \to \rho \gamma}}
\def\bromg    {\ensuremath {B \to (\rho/\omega) \gamma}}
\def\brog    {\ensuremath {\Bz \to \rho^0 \gamma}}
\def\brpg    {\ensuremath {\Bu \to \rho^+ \gamma}}
\def\bomg    {\ensuremath {\Bz \to \omega \gamma}}

\def\Kz    {\ensuremath{K^{0}}\xspace}

\def\thd     {\ensuremath {\theta_D}}

\def\cth     {\ensuremath {\cos{\theta_{H}}}}

\def\de        {\ensuremath {\Delta E^{*}}}

\def\mpl #1 #2 #3 {Mod.~Phys.~Lett.~{\bf#1},\ #2 (#3)}
\def\npb  #1 #2 #3 {Nucl.~Phys.~B~{\bf#1},\ #2 (#3)}
\def\plb  #1 #2 #3 {Phys.~Lett.~B~{\bf#1},\ #2 (#3)}
\def\pr   #1 #2 #3 {Phys.~Rep.~{\bf#1},\ #2 (#3)}
\def\prd  #1 #2 #3 {Phys.~Rev.~D~{\bf#1},\ #2 (#3)}
\def\prl  #1 #2 #3 {Phys.~Rev.~Lett.~{\bf#1},\ #2 (#3)}
\def\RMP  #1 #2 #3 {Rev.~Mod.~Phys.~{\bf#1},\ #2 (#3)}
\def\zpc  #1 #2 #3 {Z.~Phys.~C~{\bf#1},\ #2 (#3)}
\def\nim  #1 #2 #3 {Nucl.~Instrum.~Methods~{\bf#1},\ #2 (#3)}
\def\nima  #1 #2 #3 {Nucl.~Instrum.~Methods~A.{\bf#1},\ #2 (#3)}
\def\epjc #1 #2 #3 {Euro.~Phys.~Jour~{\bf#1},\ #2 (#3)}
\def\rmp #1 #2 #3 {Rev.~Mod.~Phys~{\bf#1},\ #2 (#3)}
\def\npbps #1 #2 #3 {Nucl.~Phys.~B.~proc.~suppl~{\bf#1},\ #2 (#3)}
\def\progtp #1 #2 #3 {Prog.~Theo.~Phys~{\bf#1},\ #2 (#3)}
\def\etal{{\it et al.}}

\newcommand{\BABARPubYear}    {03}
\newcommand{\BABARPubNumber}  {008}

\newcommand{\SLACPubNumber} {9938}
\newcommand{\LANLNumber} {0306038}

\def\figurebox#1#2#3{%
    \def\arg{#3}%
    \ifx\arg\empty1
    {\hfill\vbox{\hsize#2\hrule\hbox to #2{\vrule\hfill\vbox to #1{\hsize#2\vfill}\vrule}\hrule}\hfill}%
    \else
    {\hfill\epsfbox{#3}\hfill}%
    \fi}

\long\def\inst#1{\par\nobreak\kern 4pt\nobreak
    {\it #1}\par\vskip 10pt plus 3pt minus 3pt}

\begin{document}

\preprint{\babar-PUB-\BABARPubYear/\BABARPubNumber} 
\preprint{SLAC-PUB-\SLACPubNumber} 

\begin{flushleft}
\babar-PUB-\BABARPubYear/\BABARPubNumber\\
SLAC-PUB-\SLACPubNumber\\
hep-ex/\LANLNumber\\[20mm]
\end{flushleft}

\begin{flushright}
\today
\end{flushright}

\title{
\vskip 10mm
{\large \bf
Search for the Radiative Decays
$\brg$ and $\bomg$\\
\begin{center}
\vskip 10mm
The \babar\ Collaboration
\end{center}
}
}

%
\author{B.~Aubert}
\author{R.~Barate}
\author{D.~Boutigny}
\author{J.-M.~Gaillard}
\author{A.~Hicheur}
\author{Y.~Karyotakis}
\author{J.~P.~Lees}
\author{P.~Robbe}
\author{V.~Tisserand}
\author{A.~Zghiche}
\affiliation{Laboratoire de Physique des Particules, F-74941 Annecy-le-Vieux, France }
% ...
\collaboration{The \babar\ Collaboration}
\noaffiliation

\date{\today}

\begin{abstract}
A search for the exclusive radiative decays
$B\to\rho(770)\gamma$ and $\Bz\to\omega(782)\gamma$
is performed on a sample of about
\luminapprox\
$\BB$ events collected by the $\babar$ detector at
the \pep2 asymmetric-energy
$\epem$ storage ring.
No significant signal is seen in any of the channels. 
We set upper limits on the branching fractions $\BR$ of 
\hbox{$\BR(\brog)< \limitrhoo$},
$\BR(\brpg)<\limitrhop$, and 
$\BR(\bomg)< \limitomega$ at
$90\%$ confidence level (C.L.).
Using the assumption that
$\Gamma(\brg) = \Gamma(\brpg)=2\times\Gamma(\brog)$,
we find the combined limit $\BR(\brg)<\limitgeneric$, corresponding
to \BR(\brg)/\BR(\bkg) $<$ \limitratio\ at $90\%$ C.L.
\end{abstract}


\maketitle

Within the Standard Model (SM), 
the decays $\brg$ and $\bomg$ proceed 
primarily through an underlying $\bdg$ electromagnetic
``penguin'' diagram that contains
a top quark in the loop \cite{review}.
These processes are analogous to the $\bkg$ process mediated by 
the $\bsg$ transition, but with the final-state $s$-quark 
replaced by a $d$-quark, 
and the relevant element of the CKM matrix
changed from $V_{ts}$ to $V_{td}$.
There may also be contributions resulting from 
physics beyond the SM, such as supersymmetry \cite{hewett}.
Recent calculations of the branching fraction 
in the SM indicate a range
$\BR(\brpg)=(0.9-1.5)\times 10^{-6}$ \cite{SM,alivtdvtstheory}.
The range is due both to uncertainties in the value of
$V_{td}$ and to uncertainties in the calculation of the 
relevant hadronic form factors. 
The rates for \brog, \brpg, and \bomg\
are related by the quark model, such that we expect
$\Gamma(\brpg) \approx 2\times\Gamma(\brog) \approx 2\times\Gamma(\bomg)$.
Previous searches \cite{cleobellerg}
have found no evidence for these decays,
nor any other $\bdg$ processes.

The analysis uses data collected by the \babar\ detector \cite{ref:detector}
at the \pep2 asymmetric-energy
$\epem$ storage ring \cite{pep}.
The data sample consists of 
\lumi\ $\BB$ events corresponding to 
$\onlumi\invfb$ on the
$\FourS$ resonance (``on-resonance''), and $9.6\invfb$ recorded
$40\mev$ below the $\FourS$ resonance (``off-resonance''). 

The \babar\ detector
consists of five subdetectors.  
Charged particle trajectories are
measured by a combination of a five-layer silicon vertex tracker
(SVT)
and a 40-layer drift chamber (DCH)
in a 1.5-T solenoidal magnetic field.
Photons and electrons are
detected in a CsI(Tl) electromagnetic calorimeter (EMC), with 
photon energy resolution $\sigma_E / E = 0.023 (E/\gev)^{-1/4} \oplus 0.019$. 
A ring-imaging Cherenkov detector (DIRC) is used for charged-particle
identification. 
The magnetic flux return is instrumented with
resistive plate chambers to identify muons.

The decay $\brg$ is reconstructed with $\rho^0\to\pip\pim$ 
and $\rho^+\to\pip\piz$,
while $\bomg$ is reconstructed with $\omega\to\pip\pim\piz$. 
Charge-conjugate channels are implied throughout this paper.
Background high-energy photons are produced primarily  
in continuum $u$, $d$, $s$, and $c$ quark-antiquark events through
$\piz/\eta\to\gamma\gamma$ decays or via initial-state radiation.
The reconstruction uses quantities both in the laboratory and
$\FourS$ center-of-mass frames,
where the latter are denoted by an asterisk. 

The primary photon in the $B$ decay is identified as an
energy deposition in the EMC.
The deposition must meet a number of criteria 
(described in detail in our paper \cite{babarksg} on
\bkg)
that are designed to eliminate background
from charged particles,
hadronic showers, and $\piz$ and $\eta$ decays.

\begin{figure}[t]
\begin{center}
\begin{minipage}[t]{4.4cm}
\vskip 0.01in
\includegraphics[width=4.4 cm]{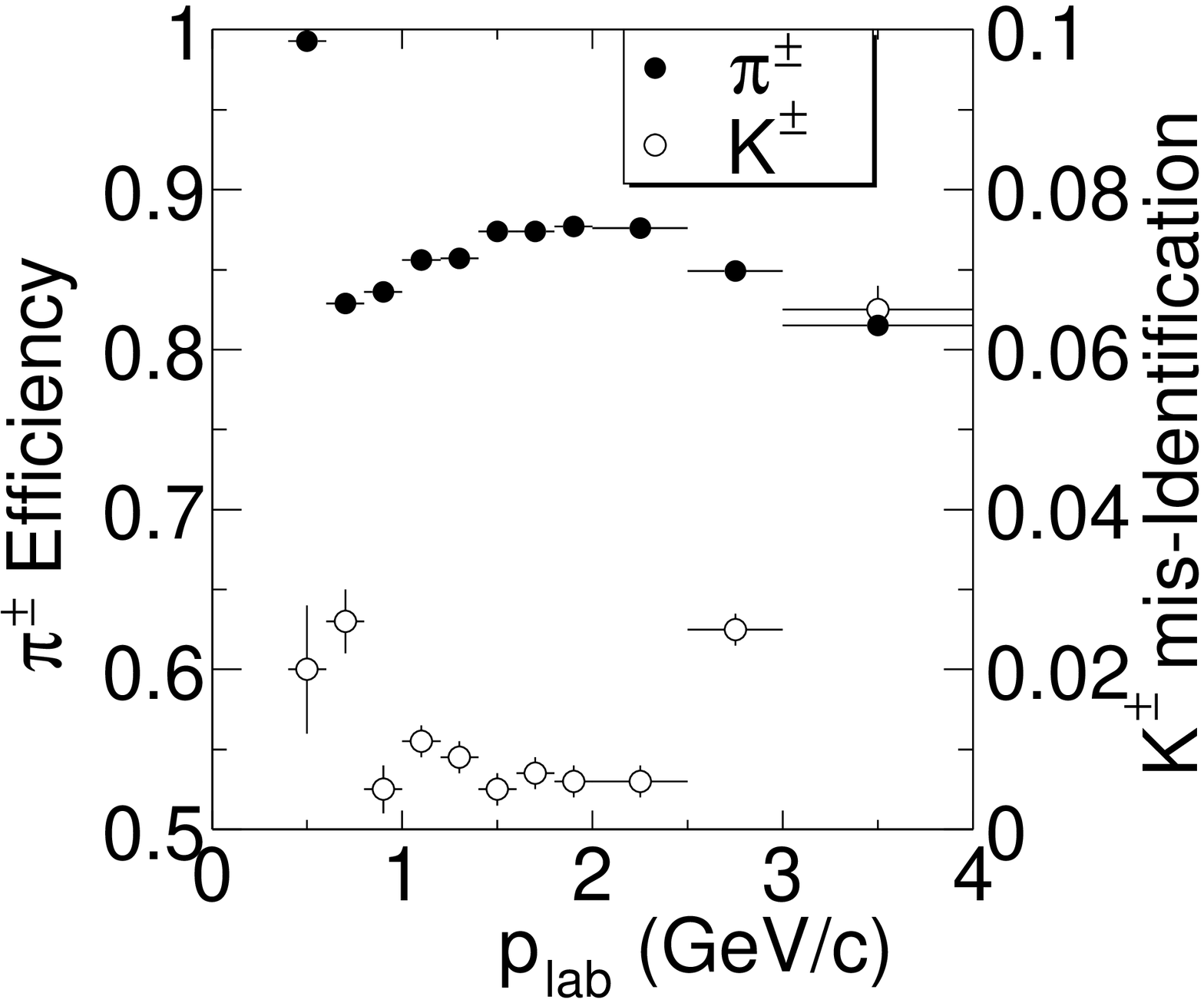}
\end{minipage}
\begin{minipage}[t]{4.1cm}
\vskip 0.07in
\includegraphics[width=4.1 cm]{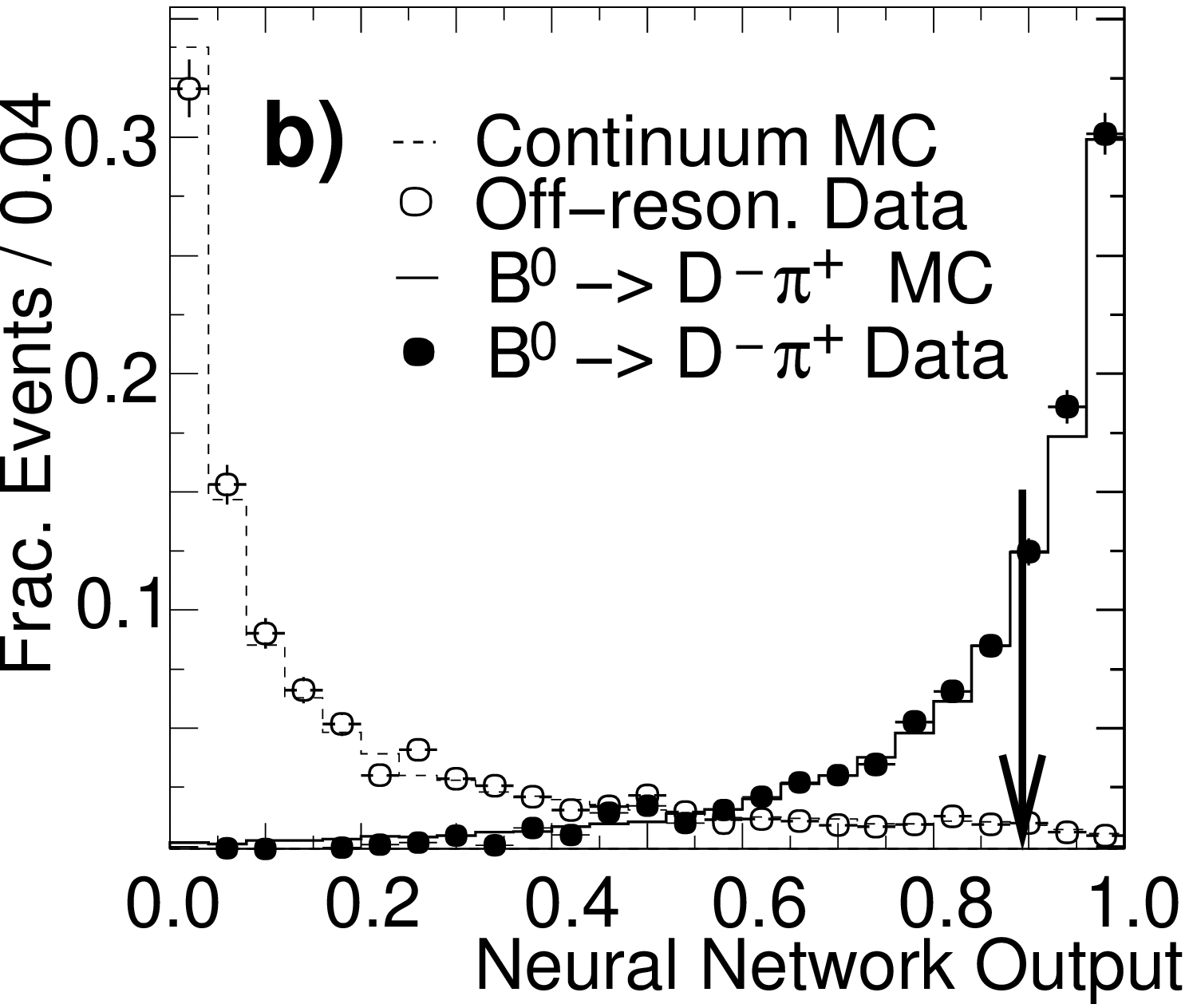}
\end{minipage}
\caption{a) Pion efficiency and kaon misidentification rate
of the charged pion selection 
for $0.4 \gevc < p_{lab} < 4.0 \gevc$, where $p_{lab}$ is 
the track momentum in the laboratory frame.
b) $\brog$ neural-network output for MC-simulated
events with comparison to data control samples. 
Events with neural-network output greater than 0.9 are selected,
as indicated.} 
\label{fig:piselectrhonn}
\end{center}
\end{figure}

As in Ref.~\cite{babarksg},
the charged tracks used in identifying the
$\rho/\omega$ meson are well-measured tracks
with a momentum transverse to the beam direction greater than 0.1 \gevc.
A charged pion selection based on $dE/dx$
measurements in the SVT and DCH,
and on Cherenkov photons reconstructed in the DIRC
is used to reduce backgrounds
from the $\bsg$ processes by rejecting charged kaons 
(e.g. $\Kp$ from $\bkog$).
Figure \ref{fig:piselectrhonn}(a) shows the particle identification 
performance measured with a control sample of 
$D^{*+}\!\rightarrow D^0(\!\rightarrow K^-\pi^+)\pi^+$ decays.

Neutral pion
candidates are identified using two photon candidates reconstructed
in the calorimeter,  each with energy greater than $50\mev$.
The invariant mass of the pair is required to satisfy
$115 < m_{\gamma\gamma} < 150\mevcc$, 
which removes pairs whose invariant mass differs from the true $m_{\pi^0}$ by
more than about 3 times the experimental resolution.
A kinematic fit with $m_{\gamma\gamma}$ constrained to 
$m_{\pi^0}$ is used to improve the momentum resolution.

A $\rho^0$ candidate is 
reconstructed by 
selecting two identified pions that have opposite charge and
a common vertex.
The $\rho^+$ candidates are obtained by pairing $\piz$
candidates with an identified charged pion.
We select $\rho$ candidates with invariant mass $m_{\pi\pi}$ 
within $250\mevcc$ of 
$m_\rho =  770\mevcc$\cite{pdg} and momentum $2.3<p_{\pi\pi}^*<2.85\gevc$.
The $\omega$ candidates are reconstructed from combinations of
oppositely charged identified pions
with a common vertex and $\piz$
candidates with invariant mass $m_{\pip\pim\piz}$ 
within $23\mevcc$ of 
$m_\omega = 783\mevcc$
\cite{pdg} and  momentum $2.4<p_{\pip\pim\piz}^*<2.8\gevc$.
The $m_{\pip\pim\piz}$ resolution is slightly poorer in 
data than in Monte Carlo (MC) simulation.
The resulting change in signal efficiency of the 
$m_{\pip\pim\piz}$ selection is accounted for
as a systematic error in the signal efficiency.

The photon and $\rho/\omega$~meson candidates are
combined to form the $B$ meson candidates.
We define
$\de \equiv E^*_{B}-E_{\rm beam}^*$,
where $E_{\rm beam}^*$ is the energy of each beam and
$E^*_B=E^*_{\gamma}+E^*_{\rho/\omega}$ is the energy
of the $B$~meson candidate. 
The signal candidates are centered at $\de=0$
with  resolution of about $50\mev$ and a tail towards negative
$\de$ due to the asymmetric energy response of the EMC.
We also define the beam-energy-substituted mass
$\mes \equiv 
\sqrt{ E^{*2}_{\rm beam}-\mbox{$\boldmath{\mathrm p'}$}_{B}^{*2}}$,
where ${\mathrm p'}_B^*$ is the momentum of the $B$ candidate
modified by scaling the photon energy to make
$E^*_\gamma+E^*_{\rho/\omega}-E_{beam}^*=0$. 
This procedure reduces the tail in the signal $\mes$ distribution 
that results from the asymmetric calorimeter response.
The signal candidates peak at $\mes=m_B$
with a resolution of about $3\mevcc$, dominated
by the beam-energy spread.

We consider candidates in the ``fit region'' $-0.3 < \de <0.3 \gev$ and
$5.20 < \mes<  5.29\gevcc$.
For the small fraction of events (8\% for MC \brog\ events)
in which more than one $B$ meson candidate satisfies all the cuts,
the candidate with the smallest value of $|\de|$ is selected.

We construct a number of variables that distinguish the signal 
from the continuum $q\bar q$ background.
As in Ref.~\cite{babarksg}, we calculate the 
thrust angle $\theta_T^*$, 
the $B$-production angle $\theta_B^*$,
and the helicity angle $\theta_H$.
For $\bomg$, $\theta_H$ is defined as 
the angle between the normal to the decay plane of the $\omega$ 
and the flight direction of the $B$ meson, both computed in the
$\omega$ rest frame.
We also calculate several additional discriminating variables.
The energy flow of the 
event excluding the $B$-meson daughters in 
$10^\circ$ cones centered on 
the photon-candidate momentum  
provides discrimination between the jet-like 
continuum background and the more spherical signal events.
For suppression of the initial-state radiation background, we 
consider $R_2'$, the ratio of second- to zeroth-order Fox-Wolfram 
moments \cite{fox} in the frame recoiling from the photon momentum.
We define the net flavor content as 
$\sum_i |N^+_i-N^-_i|$, where $N^\pm_i$ are 
the number of $e^\pm$, $\mu^{\pm}$, $K^\pm$, and slow pions of 
each sign identified in the event \cite{babarprd}.
On average, $\BB$ events have larger net flavor than continuum events.
In the $\brog$ and $\bomg$ analyses, we use the 
seperation along the beam axis of the $B$-meson candidate vertex and
that of the rest of the event.
Due to the finite $B$ lifetime, this should be larger in magnitude in
$\BB$ events than in continuum background.
In the $\bomg$ analysis, we use the $\omega$ 
Dalitz angle $\thd$, which
is defined as the angle between the $\piz$ and the $\pip$ in
the $\pip\pim$ rest frame \cite{muirhead};
$\cos{\thd}$ follows a $\sin^2\theta_D$ distribution for true $\omega$ decays,
as opposed to the uniform distribution of combinatorial background.

The background-suppression variables are combined into one discriminating
variable via a neural network, which responds
non-linearly to the input variables and exploits 
correlations between the variables \cite{snns}.
A separate neural network is trained for each mode.

The output for the neural network trained for $\brog$ is shown in
Fig.~\ref{fig:piselectrhonn}(b), where the MC
simulation of the continuum background is compared with the off-resonance 
data, and the output for MC-simulated $\Bz\to D^-\pip$ decays 
is compared with $\Bz\to D^-\pip$
decays reconstructed in the on-resonance data.
The latter comparison provides a cross-check of those input 
variables that depend on the properties of the other $B$ meson
in the event.
This includes all of the variables except for
$\theta_H$ and $\theta_D$, which, for this check, are modeled
using the signal MC distributions.

To suppress the
continuum background, we make a selection on the neural-network 
output that is optimized for minimum statistical error
as determined using MC samples of signal and background.
The efficiency of this selection for the $\B\to D\pi$
control sample differs slightly between the
data and MC.
We account for this difference as a systematic error in the 
signal efficiency.
For $\brpg$, we also require
$|\cth|<0.6$ to reject $\Bu\to\rho^+\piz$ events, which have
a $\cos^2\theta_H$ distribution,
as opposed to the expected $\sin^2\theta_H$
distribution of the signal process.

\begin{figure}[t]
\begin{center}
\includegraphics[width=6 cm]{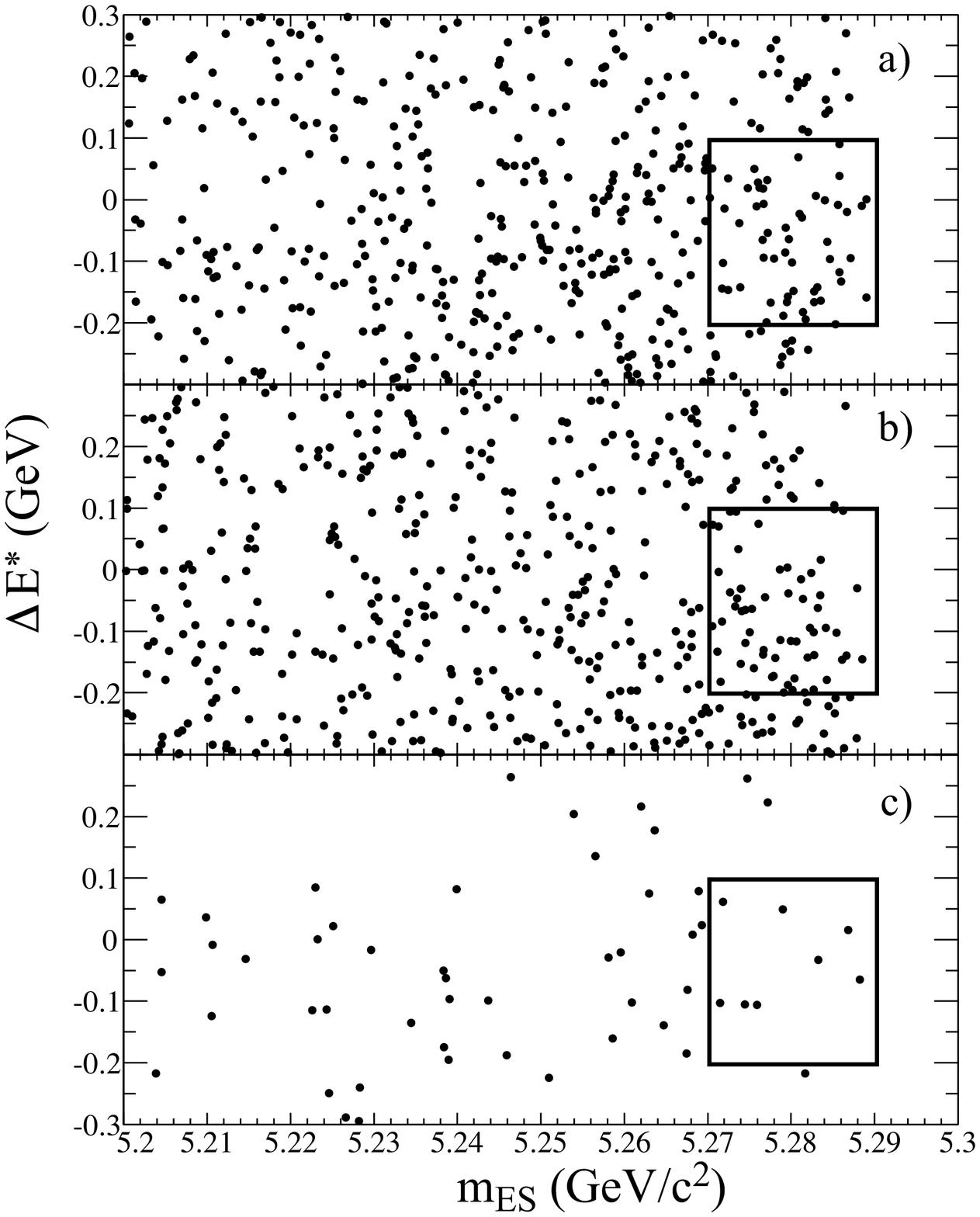} 
\caption{$\de$ vs. $\mes$ fit regions for a) $\brog$, 
b) $\brpg$, and c) $\bomg$ candidates. 
The boxes indicate the regions where signal events would appear:
$-0.2 < \de <0.1\gev$ and $5.27<\mes<5.29\gevcc$.
Assuming $\BR(\brog) =\frac{1}{2}\BR(\brpg)= \BR(\bomg) = 10^{-6}$, we 
expect 9.9, 12.1 and 3.4 signal events in these regions, respectively.
}
\label{fig:rhomesde}
\end{center}
\end{figure}

After applying the neural-network, \cth, and 
fit-region selection to the on-resonance data,
449 events remain in the $\brog$ data,
480 events for $\brpg$ and 
54 events for $\bomg$.
MC studies indicate that about 90\% of the background 
in these samples comes from continuum events, and only about
10\% from \BB.

For the signal extraction, we perform
an unbinned extended maximum likelihood fit to 
the selected events.
For $\brg$, the fit uses $\mes$, $\de$, and $m_{\pi\pi}$, 
whereas for $\bomg$, only $\mes$ and $\de$ are used.
The measured variables are largely uncorrelated,
even after the $p_{\pi\pi}^*$ (or $p_{\pi^+\pi^-\pi^0}^*$) cut, 
allowing the probability density function (PDF) to be 
constructed as a 
product of independent distributions for each variable.
Since the $\BB$ backgrounds have PDFs that largely resemble
continuum but are much smaller, 
the signal extraction uses only a continuum component
to describe the background.
Biases due to $\BB$ backgrounds are considered below.
The signal $\mes$ and $\de$ distributions
are described by the Crystal Ball shape \cite{cbshape}, 
with the exception of the $\mes$ distribution for $\brog$, 
where the Gaussian distribution is used.
The relativistic Breit-Wigner lineshape is used for the
signal $m_{\pi\pi}$ distribution.
The signal PDF parameters are obtained from MC simulation.
The background $\mes$ and $\de$ distributions are
described by the ARGUS threshold function \cite{argus} and 
a second-order polynomial, respectively.
The background $m_{\pi\pi}$ function is a sum
of a Breit-Wigner component and a combinatorial component 
described by a first order polynomial.
The background PDF parameters are determined in the fit, with
the exception of the $m_{\pi\pi}$ resonant fraction, which is
fixed to the value measured in off-resonance data.
 
The $\de$ vs. $\mes$ distributions of the selected $\brg$ and $\bomg$ 
candidates are shown in Fig.~\ref{fig:rhomesde} and the fitted 
signal yields 
are shown in Table~\ref{tab:sigyield}.
No significant signal is seen in
any mode.
The quality of the fit is checked by
comparing the overall likelihood of the fit with values
obtained from an ensemble of parameterized MC simulations
and found to be within the range expected.

\begin{table}
\begin{center} 
\begin{tabular}{lccccc} \hline\hline
Mode    & Yield   &  Bias   & Upper Lim. & $\epsilon$ & ${\cal B}$   \\
        & (Events)&  (Events)  &(Events)& $(\%)$     &  ($10^{-6}$) \\ 
\hline
$\brog$ & $4.8^{+5.7}_{-4.7}$ &[$-$0.5,0.8] &12.4& 12.3 &$0.4^{+0.6}_{-0.5}$\\
[0.05in]
$\brpg$ & $6.2^{+7.2}_{-6.2}$ &[$-$0.1,2.0] &15.4 &$\>$ 9.2 &$0.7^{+0.9}_{-0.8}$\\
[0.05in]
$\bomg$ & $0.1^{+2.7}_{-2.0}$ &[$-$0.3,0.5] &$\>\;$3.6& $\>$ 4.6& $0.0^{+0.7}_{-0.5}$\\ 
\hline\hline
\end{tabular}
\caption{ \label{tab:sigyield}
The signal yields and errors obtained from the signal extraction fit, 
the ranges of observed biases from $\BB$ backgrounds, selection
efficiencies ($\epsilon$), and 
the inferred branching fractions (${\cal B}$)
for $\brog$, $\brpg$, and $\bomg$ in the on-resonance data sample.
The ``Upper Lim.'' is a 90\% C.L. limit.
The efficiencies include the partial branching fractions for the 
$\rho/\omega$ decays considered.}
\end{center}
\end{table}

We consider three sources of systematic uncertainty in this 
analysis: 
the modeling of $\BB$ backgrounds,
the signal reconstruction efficiency, and
the fixed parameters of the PDFs used in the fit.
The first of these is ``additive'' in that it could 
result in background adding to the fitted signal yields.
The last two are ``multiplicative'' in that they affect the way
a given signal is interpreted as a branching fraction.

The effect that $\BB$ backgrounds have on the fitted signal yields
is studied in parameterized MC simulations in which
the $\BB$ background shape in the $\mes$-$\de$ plane is
modeled with both one- and two-dimensional distributions.
Also, the rates of the dominant background modes are 
varied within wide ranges.
For $\bsg$ (including $\bkg$),
the normalization is varied between zero and twice the
nominal value to conservatively account for uncertainties in 
kaon misidentification.
For $\Bu\to\rho^+\piz$ decays the branching fraction is varied 
between zero and twice the expected rate of 
$2\times 10^{-5}$ \cite{rhopiphenom}.
Much lower branching fractions are expected for 
$\Bz\to\rho^0\piz$ and $\Bz\to\omega\piz$ \cite{rhopiphenom},
so these cause negligible backgrounds.
The small biases shown in
Table \ref{tab:sigyield} confirm that the $\BB$
PDFs are similar to those of continuum background.

All signal-efficiency systematic uncertainties, 
except those related to the neural network and the $\omega$ mass,
which are described above, are estimated in Ref.~\cite{babarksg}.
The largest uncertainties, which arise from the neural net efficiencies,
are 5\%, 5\%, and 10\% for $\brog$,$\brpg$, and $\bomg$ respectively.
The $\piz$ efficiency also contributes a 5\% uncertainty to 
$\brpg$ and $\bomg$.

The fixed parameters of the signal PDFs are studied in fits to data
for the topologically and kinematically similar, but much more common,
$\bkg$ decays: $\bkog$, $\Kstarz\to\Kp\pim$ for 
$\brog$ and $\bkpg$, $\Kstarp\to\Kp\piz$ for $\brpg$ and $\bomg$.
In these fits, the signal PDF parameters are allowed to float.
The signal event yields are compared to those expected from the 
branching fractions measured in Ref.~\cite{babarksg} and 
found to agree. 

The statistical uncertainties of the PDF parameters,
one of which is the resonant fraction in the background 
$m_{\pi\pi}$ distribution,
are used as ranges within which we vary the 
parameters of the $\bromg$ fits.
The resulting variations in the fitted signal yield,
which amount to 5\% for $\brog$ and $\bomg$ and 10\% for $\brpg$,
are taken as systematic uncertainties.
The total multiplicative systematic error, including the signal efficiency 
uncertainty, is 8\% for $\brog$ and 13\% for $\brpg$ and $\bomg$.

We assume 
${\cal B}(\Upsilon (4S)\! \rightarrow  \BzBzb) = 
{\cal B}(\Upsilon (4S) \! \rightarrow  \BpBm) = 0.5$.
In calculating upper limits, we correct for bias from 
$\BB$ backgrounds by subtracting the smallest
observed bias, which is found to be negative for all three modes,
from the signal yield.
We include the effects of the multiplicative systematic 
uncertainties by using an extension \cite{conslacpub} 
of the method described in Ref.~\cite{candh}, wherein the
systematic and statistical errors are convolved.
The resulting $90\%$ C.L. 
upper limits for the branching fractions are
$\BR(\brog)<\limitrhoo$, $\BR(\brpg)<\limitrhop$, and
$\BR(\bomg)<\limitomega$.
Although no significant signals are seen, Table \ref{tab:sigyield}
shows the measured $B$ for each mode. 
For this calculation, we subtract a bias corresponding to the center
of the allowed range, treat the half-width of the range as the 
systematic error, and add systematic 
and statistical errors in quadrature.
 
We also calculate a combined limit for the generic process \brg\
by assuming $\Gamma(\brg)=\Gamma(\brpg)=2\times\Gamma(\brog)$ and
using the lifetime ratio 
$\tau_{B^+}/\tau_{B^0} = 1.083 \pm 0.017$ \cite{pdg}.
The resulting $90\%$ C.L. upper limit is 
$\BR(\brg) < \limitgeneric$.
Using the measured value of \BR(\bkg) \cite{babarksg}, 
this corresponds to a limit of
$\BR(\brg)/\BR(\bkg) < \limitratio.$ 

This limit may be used to constrain the ratio of CKM elements
$|V_{td}/V_{ts}|$ by means of the equation \cite{alivtdvtstheory}:
\[
\frac{{\cal B}(\brg)}{{\cal B}(\bkg)}=
\left| \frac{V_{td}}{V_{ts}} \right|^{2}
\left(\frac{1-m_{\rho}^{2}/M_{B}^{2}}{1-m_{K^{*}}^{2}/M_{B}^{2}}\right)^{3}
\zeta^{2} [1+\Delta R],
\]
where $\zeta$ describes the flavor-SU(3) breaking between $\rho$ and $K^*$, and
$\Delta R$ accounts for annihilation diagrams. 
$\Delta R$ is different for $\rho^0$ and $\rho^+$, but
we do not take this into account here.
Both $\zeta$ and $\Delta R$ must be taken from theory 
and there are several different \cite{alivtdvtstheory,grinpir}
values published. 
As an example, we choose the values $\zeta = 0.76 \pm 0.10$ and 
$\Delta R = 0.0 \pm 0.2$. 
We adjust both parameters down by one $\sigma$ 
and find the limit 
$|V_{td}/V_{ts}| < 0.34$ at 90\% C.L.

In conclusion, we have found no evidence for the exclusive $\bdg$ transitions
$\brg$ and $\bomg$ in \lumi\ $\BB$ decays studied with the $\babar$ detector.
The $90\%$ C.L. upper limits on the branching fractions are
significantly lower than previous values and start to restrict the range
indicated by SM predictions 
\cite{SM,alivtdvtstheory}.

\label{sec:Acknowledgments}


We are grateful for the excellent luminosity and machine conditions
provided by our \pep2\ colleagues, 
and for the substantial dedicated effort from
the computing organizations that support \babar.
The collaborating institutions wish to thank 
SLAC for its support and kind hospitality. 
This work is supported by
DOE
and NSF (USA),
NSERC (Canada),
IHEP (China),
CEA and
CNRS-IN2P3
(France),
BMBF and DFG
(Germany),
INFN (Italy),
FOM (The Netherlands),
NFR (Norway),
MIST (Russia), and
PPARC (United Kingdom). 
Individuals have received support from the 
A.~P.~Sloan Foundation, 
Research Corporation,
and Alexander von Humboldt Foundation.


\begin{thebibliography}{99}
\bibitem{cbshape}
The ``Crystal Ball'' lineshape is a modified Gaussian
distribution with a transition to a tail function on the low side:
$ f_{CB} = \exp({-\frac{(x-\mu)^2}{2\sigma^2}})$ for
$\frac{x-\mu}{\sigma} > \alpha$ and $A\times[B-(\frac{x-\mu}{\sigma})]^{-n}$
for $\frac{x-\mu}{\sigma} < \alpha$ where
$ A \equiv \left(\frac{n}{|\alpha|}\right)^n\exp({-\frac{1}{2}\alpha^2})$
and $ B \equiv \frac{n}{|\alpha|}-|\alpha|$ are defined such as to maintain
continuity of the function and its first derivative.
\end{thebibliography}
\end{document}